\documentclass[aip, amsmath, amssymb, reprint]{revtex4-1}

\usepackage{graphicx}
\usepackage{dcolumn}
\usepackage{bm}

\usepackage[T1]{fontenc}
\usepackage{mathptmx}
\usepackage{etoolbox}
\usepackage{enumitem}
\usepackage{hyperref}

\makeatletter
\def\@email#1#2{%
 \endgroup
 \patchcmd{\titleblock@produce}
  {\frontmatter@RRAPformat}
  {\frontmatter@RRAPformat{\produce@RRAP{*#1\href{mailto:#2}{#2}}}\frontmatter@RRAPformat}
  {}{}
}%
\makeatother
\begin{document}


\title[4-Squeeze QIA: Direction-Scaling Security]{Quaternary-Squeeze Quantum Identity Authentication: Direction-Scaling Security via Single-Mode Squeezed Light}
\author{Zhipeng Chen}%

\author{Haolun Tang}%

\author{Xiao-Qi Xiao}%
\email{xiaoxq@sdju.edu.cn}
\affiliation{School of Electronic and Information Engineering, Shanghai Dianji University, 201306, Shanghai, China}%

\author{Li-Hua Gong}
\affiliation{School of Electronic and Electrical Engineering, Shanghai University of Engineering Science, 201620, Shanghai, China}%

\date{\today}

\begin{abstract}
Quantum identity authentication (QIA) has emerged as a crucial technology for secure communication systems, particularly in the burgeoning era of quantum communications. This paper proposes a novel QIA protocol based on non-classical characteristics of squeezed light fields. By exploiting quantum noise reduction properties of quadrature squeezed coherent states, the protocol fundamentally thwarts eavesdropping attempts by Heisenberg-limited uncertainty constraints. The fidelity parameter for decoy states is utilized to detect spoofing attacks, and the dynamic key update mechanism fundamentally eliminates vulnerabilities caused by key reuse. Security information ratio analysis shows that the protocol is able to resist Gaussian-cloner attacks and detect eavesdropping. Moreover, the security threshold can be further enhanced with higher squeezing, allowing tunable protection levels adaptable to different threat scenarios. Compared with binary-squeezed protocols, our proposed four-direction (quaternary-dimensional) squeezing halves the eavesdropper's guessing probability and enlarges the fidelity gap by $29\%$, tightening the discrimination threshold without relying on extra hardware, thus facilitating practical implementation.
\end{abstract}

\maketitle


\section{Introduction}

In the past several decades, the quantum communication technology has experienced rapid development, and various types of quantum communication methods have been brought up such as quantum key distribution (QKD) \cite{QKD-BB84, QKD-2020}, quantum teleportation \cite{TEL-93, PAN-98}, quantum secure direct communication \cite{L02, D03, W17} and so on. With the rapid advancement of the quantum information science, particularly the tangible progress in quantum computer science from theoretical concept to practical implementation, communication security in quantum era has attracted increasing attention. To safeguard quantum communication, a series of quantum security technologies, such as quantum authentication \cite{D99, 50, 14}, quantum key agreement \cite{ZHOU-2004, 48, 49}, quantum secret comparison \cite{yang-2009, GY-2024, GL-2024, Zhou-2025} and so on, have been continuously put forward, and a theoretical system of quantum cryptography has gradually formed. Among these, Quantum Information Authentication (QIA), which verifies participant legitimacy, serves as the first barrier for ensuring communication security.

Initially, classical authentication mechanisms were directly integrated into quantum key distribution (QKD) protocols to achieve identity verification \cite{D99}. Based on quantum characteristics, Zeng et al. developed a quantum public-key authentication protocol that effectively mitigates man-in-the-middle attacks, thereby substantially improving system security \cite{Z00}. Soon after, a quantum identity authentication scheme based on entangled states is proposed, which achieves the dual functions of simultaneously distributing quantum keys and verifying user identities \cite{S01}. Advancing quantum technologies have diversified technical approaches in QIA systems, optimizing authentication schemes. \cite{ZH05,Y08, Zh09, H11, K15, Z19, Z20, C21, Z24}. Concurrently, heightened attention has been directed toward addressing security vulnerabilities in QIA implementations. In 2021, Choi et al. proposed a QIA scheme in the measurement device independent architecture to eliminate physical layer vulnerabilities at the detection end \cite{C21}.In recent years, the field of QIA has transitioned gradually from theoretical research to practical implementation in various application scenarios\cite{Q22, Y23, C23}.

From an application-oriented perspective, the reliance on either single-photon sources or entanglement sources significantly impedes the practical feasibility of QIA. To address this challenge, a QIA protocol based on light field squeeze technology, a well-established technique refined through decades of experimental development, is proposed in this paper. The quadrature squeezed coherent states distributed over four symmetric directions $\{0, \pi /2, \pi , 3\pi /2\}$ are used to encode secret information. Compare with earlier schemes that restrict squeezing to two directions, our four-basis approach yields a $29\%$ tighter security margin under identical squeezing levels. While balanced homodyne detection and decoy-state monitoring preserve implementation simplicity. The protocol dynamically updates the shared key after each authentication, eliminating reuse vulnerabilities and maintaining robustness against Gaussian-cloner attacks, as confirmed by finite-size secret-information-ratio analysis.

The structure of this paper is organized as follows. Section \ref{bk} presents the prior knowledge necessary for understanding the QIA protocol. Section \ref{qia} then elaborates in detail on the  QIA protocol based on quadrature squeezed state. Next, Section \ref{sa} conducts a security analysis of the protocol under the Gaussian-cloner attack. Finally, Section \ref{con} draws a brief conclusion.

\section{Basic Knowledge about the quadrature squeezed state}\label{bk}

\begin{figure*}[ht]
\centering

\begin{minipage}[b]{0.45\textwidth}
\centering
\includegraphics[width=0.5\linewidth]{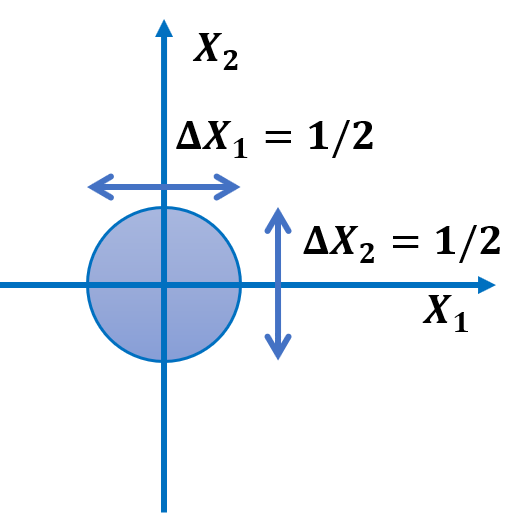}\\[-3pt]
\footnotesize (a)
\end{minipage}\hfill
\begin{minipage}[b]{0.45\textwidth}
\centering
\includegraphics[width=0.6\linewidth]{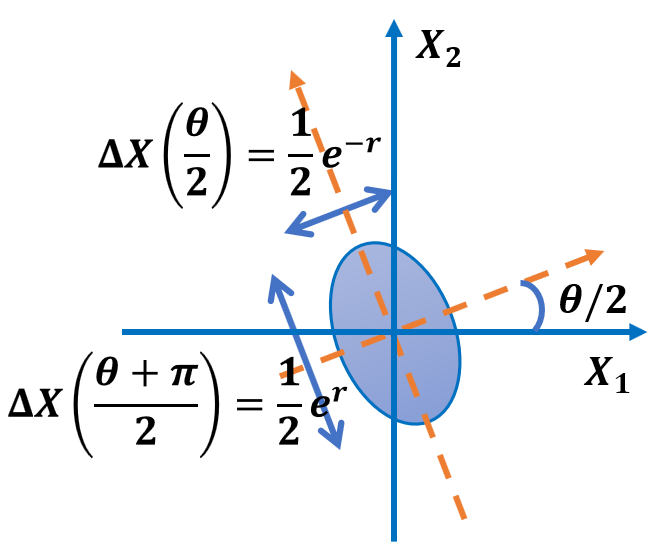}\\[-1pt]
\footnotesize (b)
\end{minipage}

\vskip 8pt

\begin{minipage}[b]{0.45\textwidth}
\centering
\includegraphics[width=0.6\linewidth]{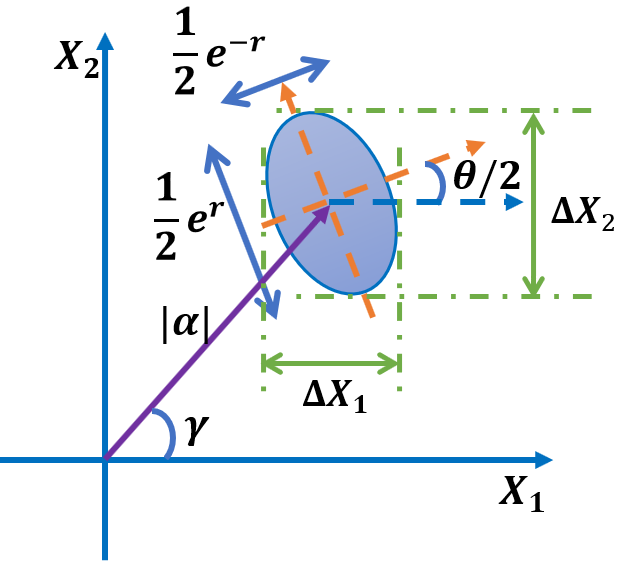}\\[-2pt]
\footnotesize (c)
\end{minipage}\hfill
\begin{minipage}[b]{0.45\textwidth}
\centering
\includegraphics[width=0.5\linewidth]{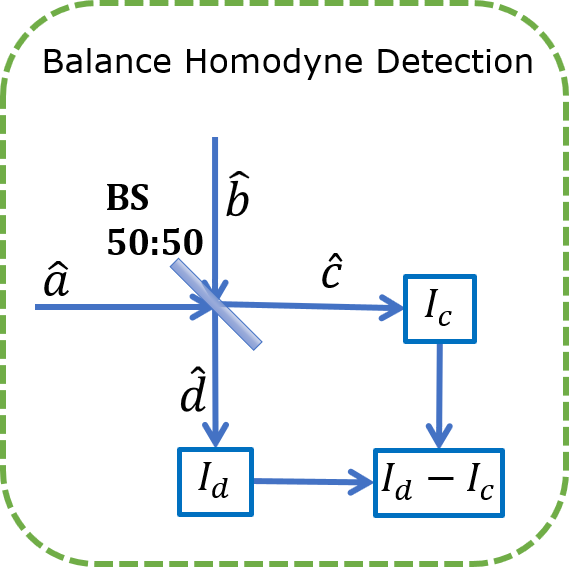}\\[3pt]
\footnotesize (d)
\end{minipage}

\caption{Phase-space pictures of (a) vacuum state, (b) squeezed vacuum state with squeezing in $X(\theta/2)$, (c) squeezed coherent state with squeezing in $X(\theta/2)$, and (d) schematic of balanced homodyne detection.}
\label{fig:total}
\end{figure*}



Before discussing the quantum identity authentication protocol, it is necessary to briefly review some basic knowledge about the quadrature squeezed state. In quantum optics, the quadrature operators of a single-mode field is defined as
\begin{equation}
\hat{X}_1=(\hat{a}+\hat{a}^\dagger)/2 \quad \mbox{ and }\quad \hat{X}_2=(\hat{a}-\hat{a}^\dagger)/(2\mathrm{i})
\end{equation}
where $\hat{a}$ and $\hat{a}^{\dagger}$ are annihilation and generation operators of the field, respectively. Evidently, $\hat{X}_1$ and $\hat{X}_2$ are associated with field amplitudes oscillating out of phase with each other by $90^{\circ }$, and satisfy uncertainty principle
\begin{equation}
\langle (\Delta \hat{X}_1)^{2}\rangle \langle (\Delta \hat{X}_2)^{2}\rangle \geq \frac{1}{16}.
\label{up}
\end{equation}
The equality in the above equation holds when the single-mode light field is in coherent state or vacuum state $\vert 0 \rangle $, and the variances of two quadratures of which are equal, $\langle (\Delta \hat{X}_1)^2\rangle = \langle (\Delta \hat{X}_2)^2 \rangle = 1/4$ at this moment, as shown in Fig. \ref{fig:total} (a) . Conveniently, a generic quadrature operator is introduced
\begin{equation}
\hat{X}(\phi )=\frac{1}{2}(\hat{a}e^{-\mathrm{i}\phi }+\hat{a}^\dagger e^{\mathrm{i}\phi }),
\end{equation}
and then there are $\hat{X}_1=\hat{X}(0)$ for $\phi = 0$, and $\hat{X}_1=\hat{X}(\pi /2)$ for $\phi = \pi /2$.

In the case of quadrature squeezing, there will be
\begin{equation}
\langle [\Delta \hat{X}(\phi )]^{2}\rangle < \frac{1}{4},
\end{equation}
which implies the fluctuations in the quadrature $X(\phi )$ are squeezed. However, the fluctuations in the other quadrature $X(\phi + \pi /2)$, orthogonal to $X(\phi )$, must be enhanced due to the uncertainty principle Eq. (\ref{up}).

To generate a quadrature squeezed state, a squeeze operator $\hat{S}(\xi )$ is applied to the single-mode field in vacuum state, for example, which is defined as
\begin{equation}
\hat{S}(\xi )= \exp [\frac{1}{2}(\xi ^{*} \hat{a}^{2} - \xi \hat{a}^{\dagger 2} )]
\end{equation}
where $\xi = re^{\mathrm{i}\theta}$ with $0 \leq r < - \infty$ is the squeeze parameter and $0 \leq \theta \leq 2\pi $ is the squeeze angle. The state of the field is $\hat{S}(\xi ) \vert 0 \rangle $, called squeezed vacuum state, and the variance of the generic quadrature of which then is
\begin{equation}
\langle [\Delta \hat{X}(\phi )]^2\rangle = \frac{1}{4}[\cosh{2r} - \sinh{2r} \cos{(\theta -2\phi )}].
\label{6}
\end{equation}
Using Eq. (\ref{6}), it is found that
\begin{equation}
\langle [\Delta \hat{X}(\frac{\theta }{2})]^2\rangle  =  \frac{1}{4}e^{-2r} \quad \mbox{and} \quad
\langle [\Delta \hat{X}(\frac{\theta + \pi }{2})]^2\rangle  =  \frac{1}{4}e^{2r},
\end{equation}
which means the maximal squeezing exists in $X(\theta /2)$, as shown in Fig. \ref{fig:total} (b). It is clear that we can get the squeezed vacuum state with the squeezing in the quadrature $X_{1}$  by choosing $\theta = 0$; and that with the squeezing in the quadrature $X_{2}$  by choosing $\theta = \pi $.

More generally, by applying a displacement operator $\hat{D}(\alpha ) = \exp(\alpha \hat{a}^{\dagger} - \alpha ^{*} \hat{a})$ to the squeezed vacuum state, a squeezed coherent state can be obtained, described as $\vert \alpha , \xi \rangle = \hat{D}(\alpha )\hat{S}(\xi)\vert 0\rangle $ with $\alpha = \vert \alpha \vert e^{\mathrm{i}\gamma }$. The phase-space representation of the squeezed coherent state is given in Fig. \ref{fig:total} (c). Evidently, there are $\alpha = X_{1} + \mathrm{i} X_{2}$, i.e., $\langle \hat{X}_{1} \rangle = \mathrm{Re}(\alpha)$ and $ \langle \hat{X}_{2} \rangle = \mathrm{Im}(\alpha)$. In this situation, the generic quadrature $X(\phi )$ is given as
\begin{equation}
\langle X(\phi )\rangle = \mathrm{Re}(\alpha e^{-\mathrm{i}\phi}),
\label{xphi}
\end{equation}
which is relative to $\alpha $; while the variance of $X(\phi )$ still takes the form in Eq. (\ref{6}). It is found that the maximum amount of quadrature squeezing still exists in $X(\theta /2)$, and in this condition there is
\begin{equation}
\langle \hat{X}(\theta /2) \rangle = \mathrm{Re}(\alpha e^{-\mathrm{i}\theta /2}),\quad \mbox{ and } \quad \langle [\Delta \hat{X}(\theta /2)]^{2} \rangle = \frac{1}{4}e^{-2r}.
\label{12}
\end{equation}

One of the methods for detecting the quadrature of the single-mode light is know as balanced homodyne detection. As shown in Fig. \ref{fig:total} (d), the light field to be detected $\hat{a}$, usually called signal light,  is mixed with a strong coherent field $\hat{b}$ (assuming be in the state $\vert \beta \rangle $ with $\beta =\vert \beta \vert e^{\mathrm{i}\psi}$), called local oscillator light, by a $50:50$ beam splitter (BS). The frequency of the local light must be the same as that of signal light. Then, the intensities of the output modes $\hat{c}$ and $\hat{d}$  of BS are measured, and the intensities difference between the two modes is
\begin{equation}
I_{c} - I_{d} = \langle \hat{c}^{\dagger}\hat{c} - \hat{d}^{\dagger}\hat{d} \rangle \propto 2\vert \beta \vert \langle \hat{X}(\phi ) \rangle
\end{equation}
where the detection angle $\phi = \psi + \pi /2$. By changing the phase $\psi $ of the local oscillator light, an arbitrary quadrature of the single-mode light field $\hat{X}(\phi ) $ can be measured.

\section{Quantum identity authentication protocol}\label{qia}

\begin{figure*}[htbp]
\centering
\includegraphics[width=0.8\textwidth]{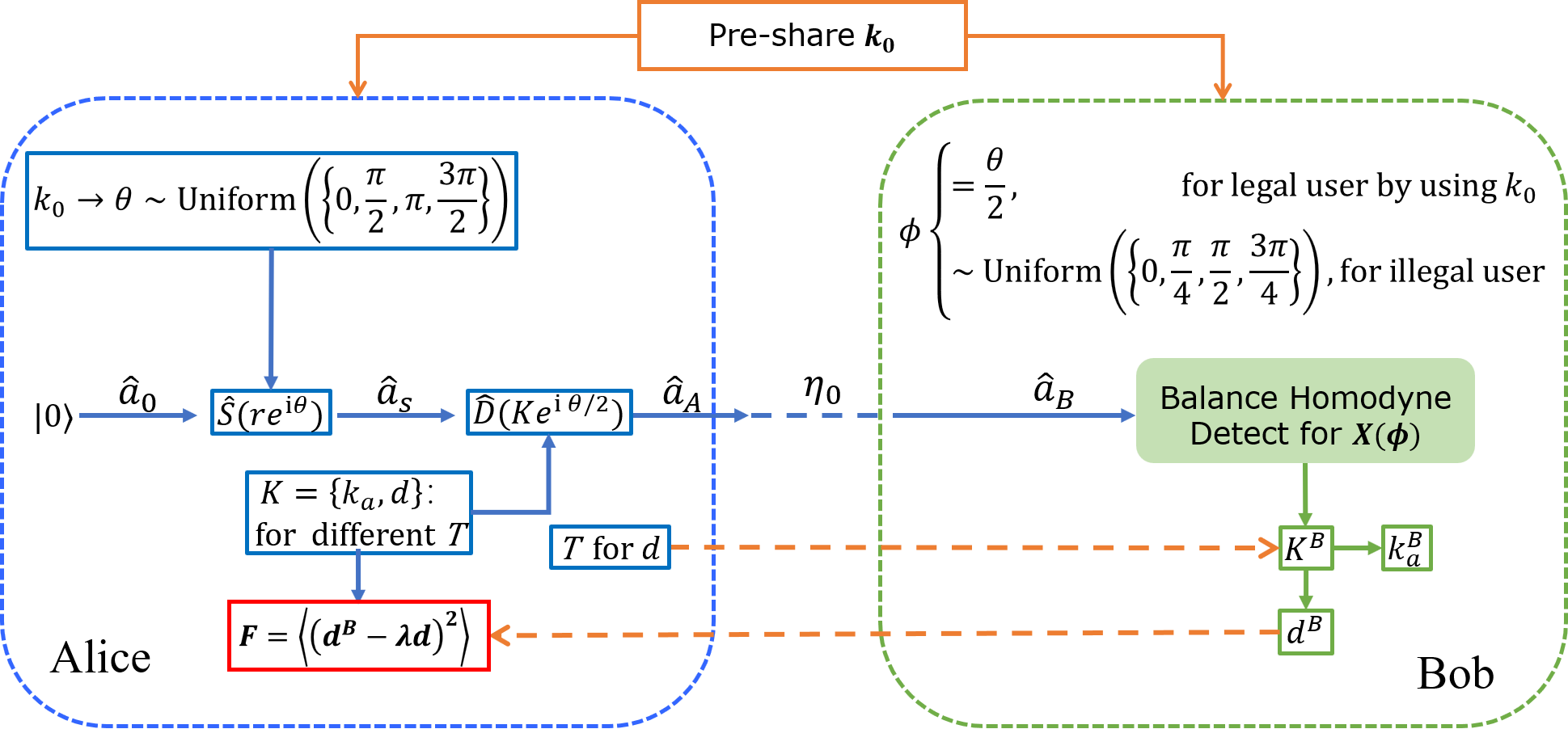}
\caption{Schematic diagram of the quantum identity authentication based on single-mode squeezed light field}
\label{fig:scheme}
\end{figure*}

A one-way quantum identity authentication protocol between two legitimate users, Alice and Bob, is described in this section. In this protocol, Alice serves as the reliable authentication server, while Bob functions as the client seeking verification. Prior to communication, Alice must authenticate Bob's identity. At the beginning, Alice and Bob share a secret key $k_{0}$ established via a secure key distribution technique (e.g., Quantum Key Distribution, QKD). The key $k_{0}$ is a set of real numbers following a Gaussian random distribution $\sim N(0, k^{2})$, with sufficiently length to satisfy the requirements of a one-time pad. The authentication workflow of the proposed protocol is schematically depicted in Fig. \ref{fig:scheme}, which comprises the following key steps.

\begin{itemize}
    \item[] \textbf{Step 1.} Alice first maps the pre-shared key $k_{0}$ to a phase parameter  $\theta $ uniformly distributed over the discrete set $\Theta =\{0, \pi /2, \pi , 3\pi /2 \}$ (via one possible implementation detailed in Appendix \ref{app:mapping}), then applies the single-mode squeezing operator $\hat{S}(re^{\mathrm{i}\theta})$ to the optical field in the vacuum state $\vert 0\rangle $.

    \item[] \textbf{Step 2.} Alice generates two Gaussian-distributed random real number sequences $k_a$ and $d$ (i.e. $k_a,d \sim N(0,k^2)$), where $k_a$ carries the secret information to be transmitted, and $d$ serves for decoy states generation. The displacement operator $\hat{D}(Ke^{i\theta/2})$ is subsequently applied to the optical field, with the amplitude $K$ randomly chosen from either sequence $k_a$ or sequence $d$ in different time slots. This process simultaneously accomplishes both secret information encoding and probabilistic decoy state insertion. The modulated optical field $\hat{a}_A$ is finally transmitted to Bob.

    \item[] \textbf{Step 3.}  Bob performs balanced homodyne detection on the optical field $\hat{a}_{B}$ he  received. Since Bob possesses the key $k_{0}$, he can employ the same mapping operator as Alice used to derive $\theta $, and then set the detection angle $\phi = \theta /2$, so that the measurement results theoretically will be $\langle \hat{X}(\theta /2) \rangle = \mathrm{Re}(Ke^{\mathrm{i}\theta /2}e^{-\mathrm{i}\theta /2}) = K$. To facilitate description, the sequence obtained from the balanced homodyne detection is denoted as $K^{B}$.

    \item[] \textbf{Step 4.}  Alice informs Bob of the time slots $T$ where $K = d$. Bob then separates the sequence $K^{B}$ into $k_{a}^{B}$ and $d^{B}$ based on $T$, and publicly broadcasts $d^{B}$.

    \item[] \textbf{Step 5.}  Alice compares the original sequence $d$ with Bob's $d^{B}$. Herein, to quantify the fidelity between sequence $d$ and $d^B$, a fidelity parameter $F$ is introduced. The fidelity parameter $F$ is defined as the minimum mean-square deviation between Bob's broadcast sequence $d^B$ and Alice's reference sequence $d$, optimized over a linear scaling factor $\lambda$:
    \begin{equation}
        F = \min_{\lambda} \left\langle (d^B - \lambda d)^2 \right\rangle
    \end{equation}
    where the coefficient $\lambda$ is chosen to minimize $d^B - \lambda d$. Therefore, Alice calculates the parameter $F$, and make a judgment whether to proceed to the next step. If the parameter $F$ is within a predetermined error margin (which is discussed below), Bob is authenticated, and the communication proceeds securely. Otherwise, the identity verification fails, and the communication is regarded as insecurity and should be broken off.

    \item[] \textbf{Step 6.}  Upon successful authentication, Alice and Bob update their shared authentication key, treating $k_{a}$ as the new key for subsequent communications.
\end{itemize}

Obviously, in the above authentication process, the threshold of the fidelity parameter $F$ plays a key role in judging of the security. Without loss of generality, we suppose the transmission coefficient of the channel is $\eta _{0}$ and the environment $\hat{a}_{v}$, which provides excess noise, is in the vacuum state $\vert 0\rangle $. In this condition, the light field Bob received can be described as
\begin{equation}
\hat{a}_{B} = \sqrt {\eta _{0}}\hat{a}_{A} +\sqrt {1-\eta _{0}}\hat{a}_{v}.
\end{equation}

If Bob is a legal user, he can obtain the squeeze angle $\theta $ from $k_{0}$ by adopting the sam algorithm as Alice used. The quadrature of the light beam Bob obtained in the case $\phi = \theta /2$ can be expressed as
\begin{equation}
\hat{X}_{B} (\theta /2)= \sqrt {\eta _{0}}\hat{X}_{A} (\theta /2)+\sqrt {1-\eta _{0}}\hat{X}_{v}(\theta /2).
\end{equation}
Thus, the sequences $d^{B}$ extracted from the measurement result of the balanced homodyne detection for $\hat{X}_{B} (\theta /2)$ at Bob's site would be $\sqrt {\eta _{0}}d$, while the variance of the sequences $d^{B}$ is $\eta _{0}k^{2} +[\eta _{0}e^{-2r} + (1-\eta _{0})]/4$, based on Eq. (\ref{12}). Using the least-squares method, the optimal $\lambda$ for the legitimate user yields
\begin{equation}
\lambda_{\mathrm{opt}} = \frac{\langle \sqrt{\eta_0}d \cdot d \rangle}{\langle d^2 \rangle} = \sqrt{\eta_0}
\end{equation}
which minimizes $d^B - \lambda d$. Thus, Alice should set $\lambda = \sqrt {\eta _{0}}$, and obtains the fidelity parameter
\begin{equation}
F_{leg} =\langle(d^{B} - \sqrt {\eta _{0}} d)^{2}\rangle _{\mathrm{min}}=\frac{1}{4} [\eta _{0}e^{-2r} + (1-\eta _{0})].
\end{equation}

Conversely, Bob is illegal, or is some eavesdropper, and gets the modified optical field from Alice.  In order to pretend to be a legal user, he/she must figure out the sequence $K$ from the light field and broadcasts $d$, as a legal user does. However, since he/she has no idea of the key $k_{0}$, he/she cannot select the measurement angle to satisfy the condition $\phi = \theta /2$. To recover $K$, the illegal user perform balanced homodyne detections with detection angles $\phi _{ill}$ randomly selected from the set $\Phi =\{0, \pi /4, \pi /2, 3\pi /4 \}$. In this situation, based on Eq. (\ref{xphi}) the sequences $d^{B}_{ill}$ obtained by the illegal user would be $\sqrt {\eta _{0}}d\cos{(\theta /2 - \phi _{ill})}$, and the variance of the sequences $d^{B}_{ill}$ would be
\begin{align}
 & \eta _{0}k^{2}\cos^{2}{(\theta /2 - \phi _{ill})} + \frac{1}{4}\bigl[\eta  _{0}(\cosh{2r} -\sinh{2r} \cos{(\theta - 2\phi _{ill})}) \notag \\
 &+ (1-\eta  _{0})\bigr],
 \end{align}
 according to Equation (\ref{6}). Therefore the fidelity parameter Alice makes out would be
\begin{eqnarray}
F_{ill} &= &\langle(d^{B} _{ill} - \sqrt {\eta _{0}} d)^{2}\rangle _{\mathrm{min}} \nonumber\\
&=& \eta _{0}k^{2}[\cos{(\theta /2 - \phi _{ill})} - 1]^{2}+ \frac{1}{4}\bigl[\eta  _{0}(\cosh{2r} \nonumber\\
&& -\sinh{2r} \cos{(\theta - 2\phi _{ill})})+ (1-\eta  _{0})\bigr],
\label{15}
\end{eqnarray}
with the predefined $\lambda = \sqrt {\eta  _{0}}$.

It is noticed that $\theta $ and $\phi _{ill}$ are selected from the uniform discrete distribution over the sets $\Theta $ and $\Phi $, respectively, and $\theta $ and $\phi _{ill}$ are independent of each other, so there are a total of $4 \times 4=16$ equally probable combinations, with a probability of $1/16$ for each combination. By enumerating all combinations and calculating, we obtain the joint distribution of $\cos{(\theta/2 - \phi _{ill})}$ and $\cos{(\theta - 2\phi _{ill})}$ which is described in table \ref{table}. Considering all four possible value sets of $\cos{(\theta/2 - \phi _{ill})}$ and $\cos{(\theta - 2\phi _{ill})}$, we can get the fidelity parameter $F_{ill}$ Alice obtained in this condition as follows (detailed derivation is provided in Appendix ~\ref{app:derivation_fidelity})
\begin{equation}
F_{ill\_ave} =\frac{1}{4}(4-\sqrt{2})\eta _{0}k^{2}+\frac{1}{4} [\eta  _{0}\cosh{2r} + (1-\eta  _{0})].
\label{16}
\end{equation}

\begin{table}[htbp]
\caption{\label{table}The possible values for $\cos{(\theta /2 - \phi _{ill})}$ and $\cos{(\theta - 2\phi _{ill})}$ and corresponding probability}
\begin{ruledtabular}
\begin{tabular}{lc}
Probability & Possible Values \\ \hline
$1/4$ & $\cos{(\theta /2 - \phi _{ill})}  =1$ and $\cos{(\theta - 2\phi _{ill})} = 1$\\
$3/8$ & $\cos{(\theta /2- \phi _{ill})} = \frac{\sqrt{2}}{2}$ and $\cos{(\theta - 2\phi _{ill})} = 0$\\
$1/4$ & $\cos{(\theta /2- \phi _{ill})} = 0$ and $\cos{(\theta - 2\phi _{ill})} = -1$\\
 $1/8$ & $\cos{(\theta /2- \phi _{ill})} = -\frac{\sqrt{2}}{2}$ and $\cos{(\theta - 2\phi _{ill})} = 0$\\
\end{tabular}
\end{ruledtabular}
\end{table}

Mathematically, when $r>0$, there always be
\begin{align}
    \Delta F &= F_{ill_ave} - F_{leg} = (1-\frac{\sqrt{2}}{4})\eta _{0}k^{2}+ \frac{\eta   _{0}}{4} (\cosh{2r} - e^{-2r}) \nonumber \\
    & >(1-\frac{\sqrt{2}}{4})\eta _{0}k^{2} > 0,
\end{align}
which implies that there is always $F_{ill} > F_{leg}$, and the difference between  $F_{ill}$ and $F_{leg}$, i.e., fidelity difference $\Delta F$ will become greater with the increasing of the squeeze parameter $r$. Therefore, Alice can set the threshold of the parameter $F$ to be $F_{leg}$ to detect illegitimate users. In other words, we can effectively discriminate the legitimacy of the user using fidelity parameters $F$, thereby thwarting the eavesdropper's spoofing attack.

Compared with earlier squeezed-light schemes that lock angles to $\theta = 0 \text{ or } \phi $ and let the eavesdropper pick $\phi _{ill}\in \{0, \pi /2\}$, the present protocol doubles the basis set to four symmetric directions. Repeating the fidelity analysis for the old binary case gives
\begin{align}
\Delta F^{(2)} = F_{\mathrm{ill\_ave}}^{(2)} - F_{\mathrm{leg}}^{(2)}
& = \frac{1}{2}\eta_{0}k^{2} + \frac{\eta_{0}}{4}(\cosh 2r - e^{-2r}) \notag\\
&> \frac{1}{2}\eta_{0}k^{2} > 0,
\end{align}
whereas the new quaternary basis yields $\Delta F > \left(1-\sqrt{2}/4\right)\eta _{0}k^{2}$, implying an immediate $29 \% $ enlargement. Thus, doubling the squeeze directions converts the former ``binary-guess" weakness into a quaternary-guess advantage: Eve's angle-matching probability drops from 1/2 to 1/4, and the enlarged fidelity gap detects illegal users earlier without any additional hardware. The same direction-scaling principle can be extended to $ 2^m $ bases for even steeper security margins.

\section{Security analysis}\label{sa}

As discussed in the above section, an illegal user cannot have any chance to successfully pass authentication unless she obtains all authentication keys, because any incorrect key will cause fidelity alterations leading to authentication failure. For  the eavesdropper Eve, whose ultimate objective is to pass identity authentication, the optimal strategy is to steal the updated authentication keys $k_{a}$ to gain access in subsequent authentication rounds. A possible attack method for Eve to steal secret information is a Gaussian-cloner attack strategy. For the convenience of discussion, in this section the transmission coefficient of the channel is assumed to be lossless, i.e., $\eta _{0} = 1$.

\subsection{Gaussian-cloner attack strategy of Eve}

\begin{figure}[htb]
\centering
\includegraphics[width=0.4\textwidth]{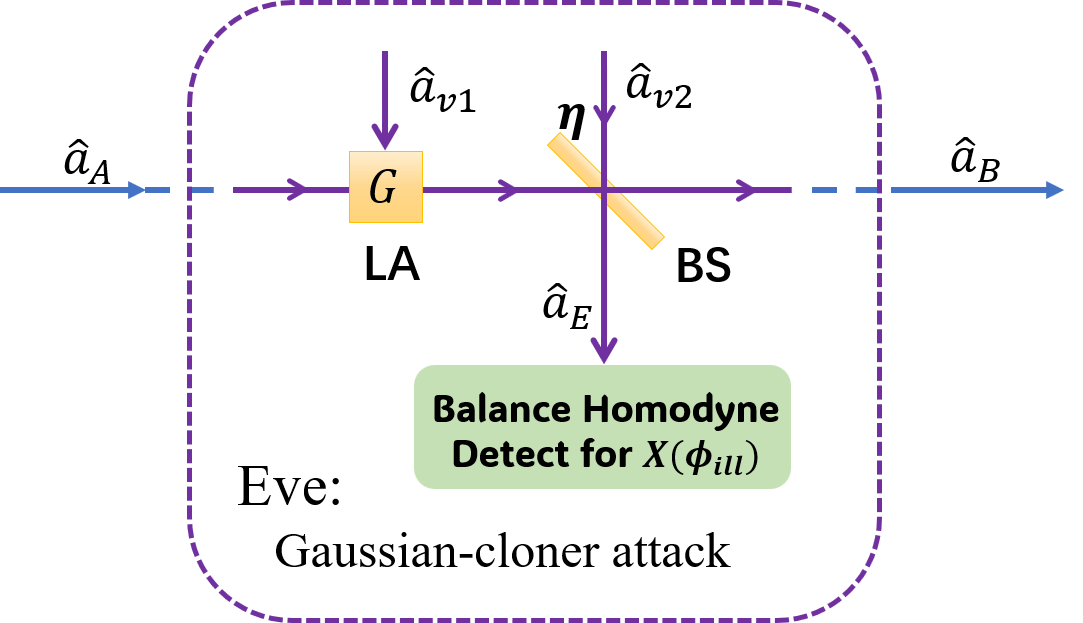}
\caption{Schematic diagram of Gaussian-cloner attack performed by Eve.}
\label{fig:gattack}
\end{figure}

In order to acquire the authentication keys, Eve amplifies the light field coming from sender Alice by utilizing a linear amplifier (LA) with gain $G$. As a result, the light beam out of LA can be expressed as $ \sqrt{G}\hat{a}_{A} + \sqrt{(G-1)}\hat{a}_{v1}^{\dagger} $, where $\hat{a}_{v1}$ represents of the environment mode. Then, the light field is divided into two beams $\hat{a}_{B} $ and $\hat{a}_{E} $ by a BS with the transmission coefficient $\eta $, as described in Figure \ref{fig:gattack}. One of the beams is sent to the receiver Bob, which takes the form
\begin{equation}
\hat{a}_{B}^{G} = \sqrt{G\eta }\hat{a}_{A} + \sqrt{(G-1)\eta }\hat{a}_{v1}^{\dagger} +\sqrt{1-\eta }\hat{a}_{v2},
\end{equation}
with $\hat{a}_{v2}$ to be the environment mode introduced by BS. And the other one $\hat{a}_{E} $ is kept by Eve, which is given by
\begin{equation}
\hat{a}_{E} = \sqrt{\eta }\hat{a}_{v2} - \sqrt{G(1-\eta )}\hat{a}_{A}+\sqrt{(1-\eta )(G-1)}\hat{a}_{v1}^{\dagger}
\end{equation}
Obviously, the Gaussian-cloner attack will be reduced to a BS attack by setting $G=1$.

\subsection{Secret information ratio}


\begin{figure*}[t]
\centering

\begin{minipage}[b]{0.48\textwidth}
\centering
\includegraphics[width=0.99\linewidth]{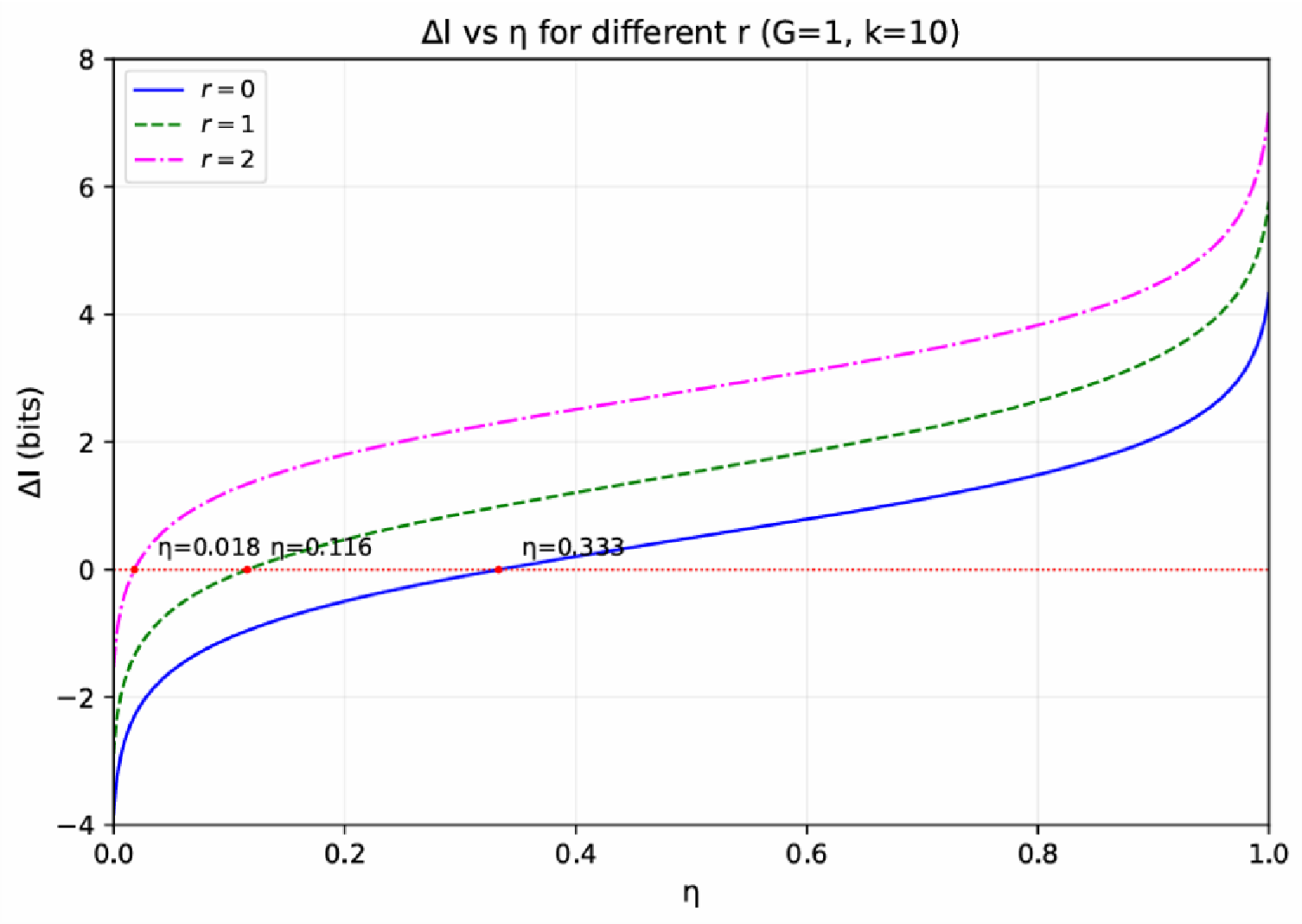}\\[-3pt]
\footnotesize (a)
\end{minipage}\hfill
\begin{minipage}[b]{0.48\textwidth}
\centering
\includegraphics[width=0.99\linewidth]{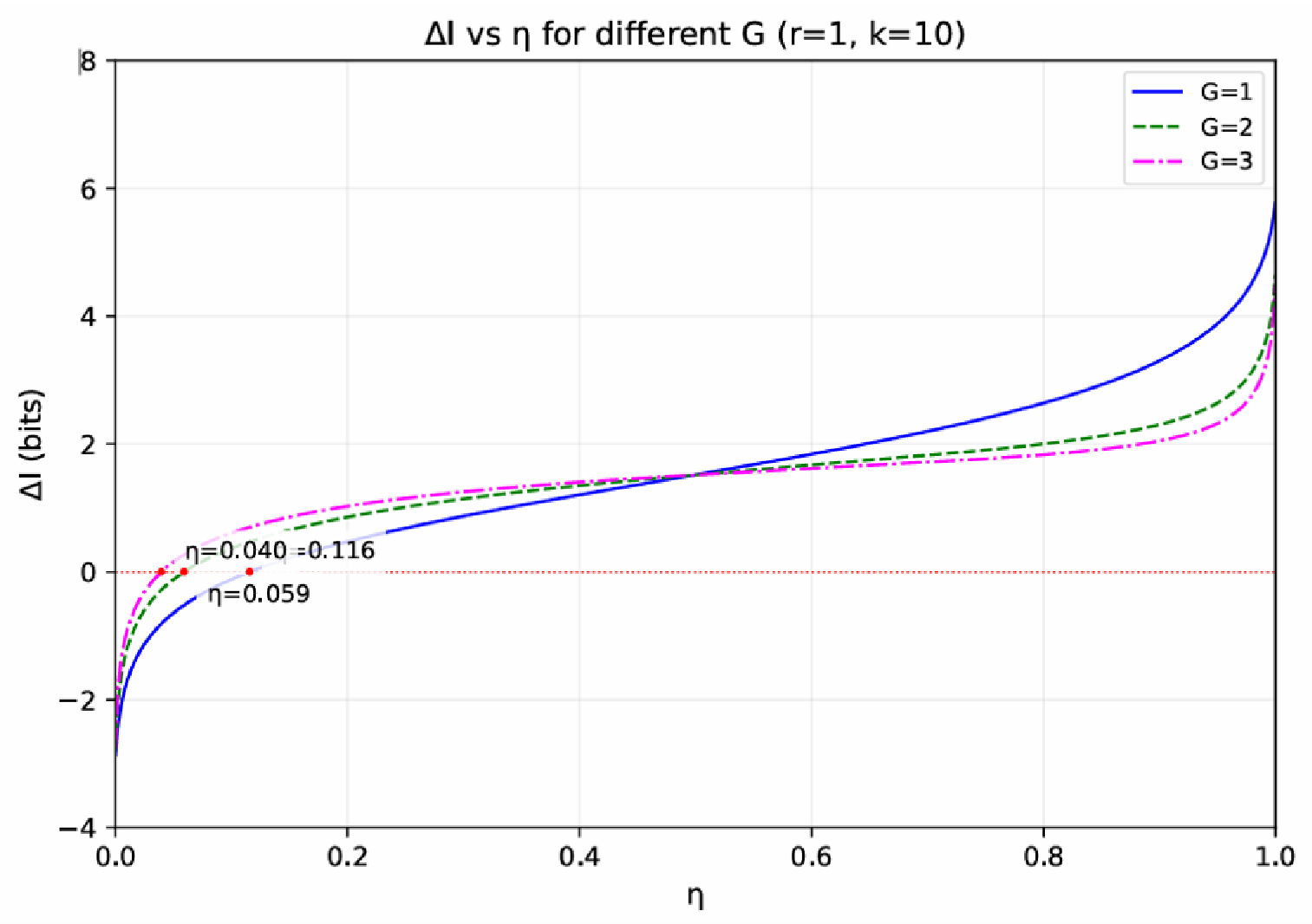}\\[-1pt]
\footnotesize (b)
\end{minipage}

\caption{The correlation between the secret information ratio $\Delta I$ and the transmission efficiency $\eta$ for (a) different squeezing parameters $r$ ($G=1$, $k=10$) and (b) different amplifier gains $G$ ($r=1$, $k=10$).}
\label{fig2:master}
\end{figure*}

In the case that the eavesdropper Eve implements a Gaussian-cloner attack, the quadrature $\hat{X}_{B}(\theta /2)$ detected by the legal user Bob can be expressed as
\begin{eqnarray}
\hat{X}_{B} (\theta /2)& = & \sqrt{G\eta }\hat{X}_{A} (\theta /2)+ \sqrt{(G-1)\eta }\hat{X}_{v1}^{\dagger}(\theta /2) \nonumber  \\
&&+\sqrt{1-\eta }\hat{X}_{v2}(\theta /2).
\end{eqnarray}
Assuming the environment modes $\hat{a}_{v1}$ and $\hat{a}_{v2}$ to be in vacuum state, the variance of $\hat{X}_{B}(\theta /2)$ then is given as
\begin{equation}
\langle [\Delta \hat{X}_{B}(\theta /2)]^{2} \rangle = G\eta k^{2} + \frac{1}{4}[G\eta e^{-2r} + (G-1)\eta +(1-\eta )].
\end{equation}
Therefore, we can get the signal-to-noise ratio of the quantum channel from Alice to Bob
\begin{equation}
\mathrm{SNR}_{AB}= \frac{4G\eta k^{2}}{G\eta e^{-2r} + (G-1)\eta +(1-\eta )},
\label{snrab}
\end{equation}
and the mutual information between Alice and Bob is in the form
\begin{equation}
I_{AB}=\frac{1}{2}\log_{2}{(1+\mathrm{SNR}_{AB})}.
\label{iab}
\end{equation}
As expected, the mutual information $I_{AB}$ reduces to $\log_{2}{(1+4k^{2}/e^{-2r})}/2$ when Eve does not exist, i.e., $G = \eta =1$. At this moment, the mutual information $I_{AB}$ is just the capacity of the channel between Alice and Bob, which increases as $r$ and $k^{2}$ increase.

On the other hand, in order to extract the information of $k_{a}$ from the light beam $\hat{a}_{E}$, Eve carries out balanced homodyne detections to measure the quadrature $\hat{X}(\phi _{ill})$, as described in the above section, which has the form
\begin{eqnarray}
\hat{X}_{E}(\phi _{ill}) &=& \sqrt{\eta }\hat{X}_{v2}(\phi _{ill}) - \sqrt{G(1-\eta )}\hat{X}_{A}(\phi _{ill}) \nonumber \\
&&+\sqrt{(1-\eta )(G-1)}\hat{X}_{v2}^{\dagger}(\phi _{ill}).
\end{eqnarray}
Since modes $\hat{a}_{v1}$ and $\hat{a}_{v2}$ are in vacuum state, the variances of $\hat{X}_{E}(\phi _{ill})$ is given as
\begin{eqnarray}
\langle [\Delta \hat{X}_{E}(\phi _{ill})]^{2} \rangle & = & G(1-\eta )k^{2}\cos^{2}{(\theta  /2 - \phi _{ill})} \nonumber \\
&&+ \frac{1}{4}\{G(1-\eta )[\cosh{2r}-\sinh{2r}\cos{(\theta -2\phi _{ill})}]\nonumber \\
&&+ (G-1)(1-\eta )+\eta \}
\end{eqnarray}
As listed in table \ref{table} there are four sets of possible values of $\cos{(\theta/2 - \phi _{ill})}$ and $\cos{(\theta - 2\phi _{ill})}$. Therefore, taking the four possible cases into consider, in whole the variances of $\hat{X}_{E}(\phi _{ill})$ is
\begin{eqnarray}
\langle [\Delta \hat{X}_{E}(\phi _{ill})]^{2} \rangle  &=& \frac{1}{2}G(1-\eta )k^{2}
+ \frac{1}{4}[G(1-\eta )\cosh{2r} \nonumber \\
&&+ (G-1)(1-\eta )+\eta ].
\end{eqnarray}
Thus, the signal-to-noise ratio in communication between Alice and Eve can be expressed as
\begin{equation}
\mathrm{SNR}_{AE}  =  \frac{2G(1-\eta )k^{2}}{G(1-\eta )\cosh{2r}+ (G-1)(1-\eta )+\eta },
\label{snrae}
\end{equation}
and the mutual information between Alice and Eve is
\begin{equation}
I_{AE}=\frac{1}{2}\log_{2}{(1+\mathrm{SNR}_{AE})}
\label{iae}
\end{equation}

The security of the proposed quantum communication protocol is quantified by the secret information ratio $\Delta I = I_{AB} - I_{AE}$. The protocol maintains unconditional security when $\Delta I > 0$, indicating legitimate communication is protected against eavesdropping. As derived from Eqs. (\ref{snrab})-(\ref{iae}), the secret information ratio exhibits strong dependence on the transmission efficiency $\eta$. Fig. \ref{fig2:master} demonstrates that $\Delta I$ monotonically decreases with reducing $\eta$. Notably, Fig. \ref{fig2:master} (a) reveals a critical threshold: when the squeezing parameter $r=0$, the system becomes vulnerable ($\Delta I \leq 0$) at $\eta < 1/3$, allowing Eve to compromise the secret key $k_{a}$.

However, this security boundary can be significantly extended by increasing the squeezing parameter $r$. Our analysis shows that higher $r$ values require Eve to achieve substantially lower $\eta$ to breach the system ($\Delta I \leq 0$). Furthermore, as evidenced by Fig. \ref{fig2:master} (b), this security enhancement exhibits robustness against Eve's amplification attempts - any increase in her gain $G$ paradoxically necessitates an even more stringent reduction in $\eta $ to compromise the security of communication between Alice and Bob ($\Delta I \leq 0$). This establishes an important design principle: Alice can enhance communication security by operating the protocol with larger squeezing parameters, thereby creating a wider security margin against potential attacks.

\subsection{Detection of Eve}

The preceding analysis of the secret information ratio $\Delta I$ indicates that Eve could potentially acquire sufficient information from the light beam $\hat{a}_{E}$ to compromise the secret key $k_{a}$ by manipulating the transmission efficiency $\eta$ beneath $1/3$ or to be much lower. However, such intrusive intervention would inevitably trigger the security monitoring system.

\begin{figure}[htbp]
\centering
\includegraphics[width=0.45\textwidth]{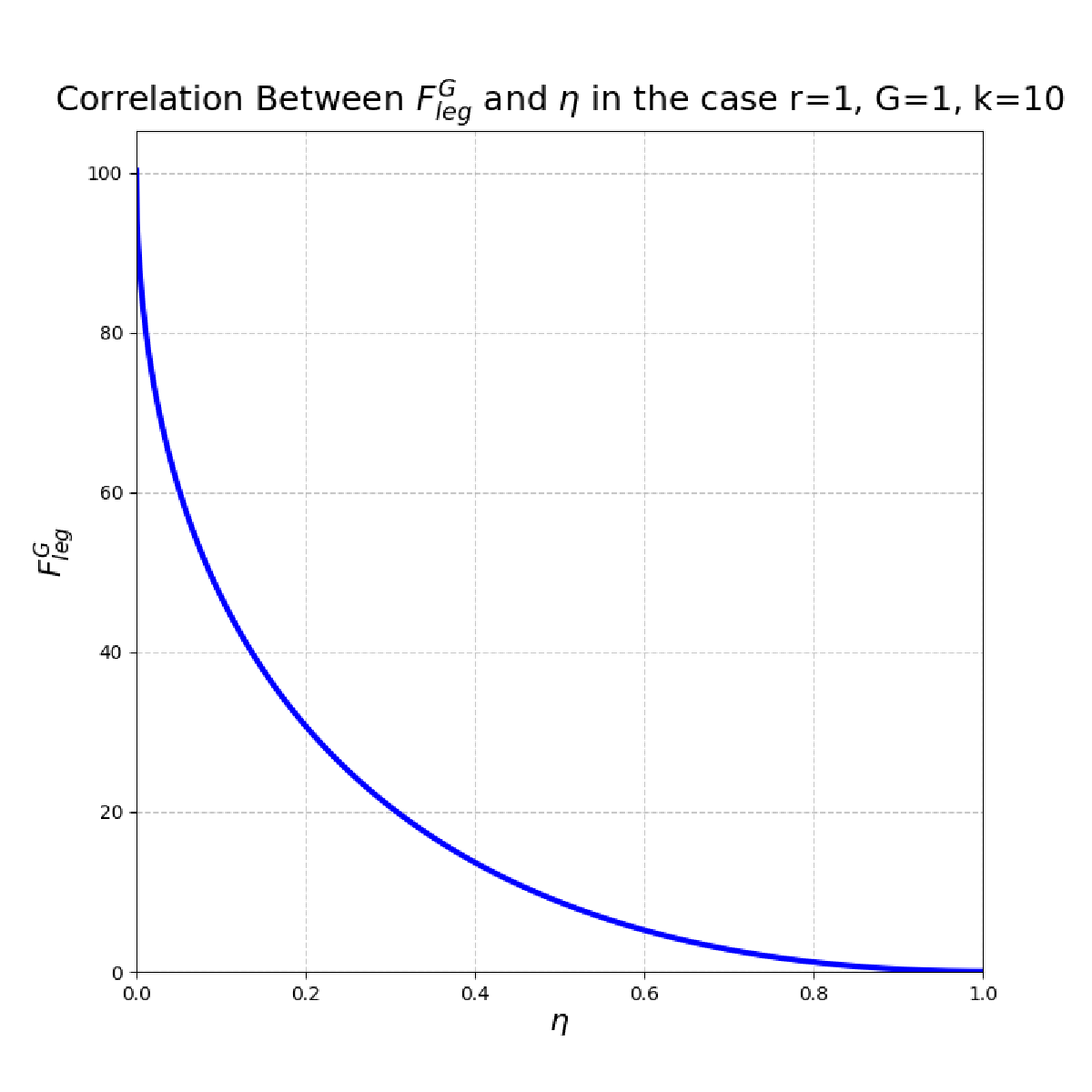}
\caption{The secret information ratio $\Delta I $ varies with $\eta $.}
\label{fig3}
\end{figure}

As specified in Step 4 of the protocol, Alice systematically evaluates the fidelity parameter $F$ to assess channel integrity. This provides a sensitive quantum fingerprint for detecting potential eavesdropping activities. Specifically, when Eve employs Gaussian cloning attacks, the fidelity parameter between the two legitimate parties has the form:
\begin{equation}
F_{leg}^{G}=(\sqrt{G\eta} -1)^{2}k^{2}+\frac{1}{4}[G\eta e^{-2r} + (G-1)\eta +(1-\eta )].
\end{equation}
The above equation reveals the critical dependence of $F_{leg}^{G}$ on the transmission efficiency $\eta $. As demonstrated in Figure \ref{fig3}, the fidelity parameter exhibits nonlinear enhancement with decreasing $\eta $, showing particularly pronounced sensitivity in the regime $\eta <1/3$. This establishes an important security feature: any attempt by Eve to intercept the secret key $k_{a}$ through reducing $\eta $ inevitably induces detectable quantum state disturbance. The resulting fidelity deviation serves as an unambiguous signature of eavesdropping, enabling Alice to abort compromised transmissions.

\section{Conclusion}\label{con}

In summary, a novel quantum identity authentication (QIA) protocol  based on single-mode squeezed light fields has been presented. By leveraging quadrature squeezed coherent states and pre-shared secret keys, the protocol establishes a robust security against sophisticated eavesdropping attempts. The fidelity parameter $F$ of the decoy states serves as a reliable metric for identifying eavesdropping attempts and distinguishing legitimate users. The security analysis reveals that  the security margin of the protocol is adjustable via squeezing parameter optimization. Dynamic key updating mechanism after authentication can efficiently prevent vulnerabilities of key reusing. The protocol requires only two practical quantum technologies for implementation: light field quadrature squeezing, and balanced homodyne detection. This minimal technological requirement ensures both practical feasibility and seamless compatibility with existing continuous-variable quantum communication systems.
 The central advance is a four-direction squeezing basis that reduces Eve's angle-guessing probability from 1/2 to 1/4 and enlarges the fidelity gap $\Delta F$ by $29 \%$, yielding a steeper security margin without additional hardware. This direction-scaling approach is readily extendable to $2^m$ bases for even stronger protection. Therefore, we believe the proposed work significantly advances the practical implementation of quantum authentication, providing a secure and technically feasible solution for emerging quantum communication infrastructures.

\begin{acknowledgments}
The authors acknowledge support from the National Natural Science Foundation of China under Grant No. 62161025, and the Program of Aboard Learning for Faculty Development in Universities of Shanghai China.
\end{acknowledgments}

\section*{AUTHOR DECLARATIONS}

\subsection{Conflict of Interest}
The authors have no conflicts to disclose.

\subsection{Author Contributions}
Zhipeng Chen: Formal analysis (equal); Project administration (equal); Writing - original draft (equal). Haolun Tang: Investigation (equal);Validation(equal); Visualization(equal).Xiao-Qi Xiao: Conceptualization (equal); Supervision (equal); Writing - review \& editing (equal). Li-Hua Gong: Supervision (equal); Funding acquisition

\section*{DATA AVAILABILITY}
Data sharing is not applicable to this article as no new data were created or analyzed in this study.

\appendix
\section{Detailed Derivation of Average Illegitimate-User Fidelity}
\label{app:derivation_fidelity}

This appendix presents a step-by-step, reproducible derivation of the average illegitimate-user fidelity $F_{ill\_ave}$ (Eq.~(\ref{16}) in the main text). The derivation enumerates all discrete angle combinations between the legitimate squeeze angles and the eavesdropper's detection angles, then groups these combinations by their cosine value pairs to compute the weighted average fidelity.

\subsection{Enumeration of Angle Combinations}

\begin{table}[htbp]
  \centering
  \caption{Complete Enumeration of 16 Angle Combinations and Corresponding Cosine Terms}
  \label{tab:angle_combinations}
  \begin{tabular}{c c c c c c}
    \toprule
    No. & $\theta$ (rad) & $\phi_{\text{ill}}$ (rad) & $c_1$ & $c_2$ & Probability \\
    \hline
    1  & 0              & 0                        & 1                                                      & 1                                              & 1/16        \\
    2  & 0              & $\pi/4$                  & $\sqrt{2}/2$                                           & 0                                              & 1/16        \\
    3  & 0              & $\pi/2$                  & 0                                                      & $-1$                                            & 1/16        \\
    4  & 0              & $3\pi/4$                 & $-\sqrt{2}/2$                                          & 0                                              & 1/16        \\
    5  & $\pi/2$        & 0                        & $\sqrt{2}/2$                                           & 0                                              & 1/16        \\
    6  & $\pi/2$        & $\pi/4$                  & 1                                                      & 1                                              & 1/16        \\
    7  & $\pi/2$        & $\pi/2$                  & $\sqrt{2}/2$                                           & 0                                              & 1/16        \\
    8  & $\pi/2$        & $3\pi/4$                 & 0                                                      & $-1$                                            & 1/16        \\
    9  & $\pi$          & 0                        & 0                                                      & $-1$                                            & 1/16        \\
    10 & $\pi$          & $\pi/4$                  & $\sqrt{2}/2$                                           & 0                                              & 1/16        \\
    11 & $\pi$          & $\pi/2$                  & 1                                                      & 1                                              & 1/16        \\
    12 & $\pi$          & $3\pi/4$                 & $\sqrt{2}/2$                                           & 0                                              & 1/16        \\
    13 & $3\pi/2$       & 0                        & $-\sqrt{2}/2$                                          & 0                                              & 1/16        \\
    14 & $3\pi/2$       & $\pi/4$                  & 0                                                      & $-1$                                            & 1/16        \\
    15 & $3\pi/2$       & $\pi/2$                  & $\sqrt{2}/2$                                           & 0                                              & 1/16        \\
    16 & $3\pi/2$       & $3\pi/4$                 & 1                                                      & 1                                              & 1/16        \\
    \hline
    \hline
  \end{tabular}
\end{table}

The proposed quantum identity authentication (QIA) protocol relies on two discrete angle sets, defined consistently with Section III:
\begin{itemize}
    \item Legitimate squeeze angles (Alice's phase set)
    \begin{equation}
        \Theta = \{0, \pi/2, \pi, 3\pi/2\},
    \end{equation}
    where each $\theta \in \Theta$ corresponds to the phase of the single-mode squeezed light field (determined by the pre-shared Gaussian key $k_0 \sim \mathcal{N}(0, k^2)$.

    \item Eavesdropper's detection angles (Eve's phase set)
    \begin{equation}
        \Phi = \{0, \pi/4, \pi/2, 3\pi/4\}
    \end{equation}
    where Eve randomly selects $\phi_{ill} \in \Phi$ to perform balanced homodyne detection (since she lacks the pre-shared key $k_0$ to match Alice's squeeze angle).
\end{itemize}

Since $\theta$ and $\phi_{\text{ill}}$ are selected independently and uniformly from their respective sets, there are $4 \times 4 = 16$ equally probable angle combinations (each with probability $1/16$). All combinations, along with their corresponding cosine terms $c_1 = \cos\left(\theta/2 - \phi_{ill}\right)$ and $c_2=\cos\left(\theta - 2\phi_{ill}\right)$ are listed in Table~\ref{tab:angle_combinations}.

\subsection{Grouping by Cosine Value Pairs}

To simplify the average fidelity calculation, the 16 combinations in Table ~\ref{tab:angle_combinations} are grouped into four distinct cases based on identical $(c_1, c_2)$ pairs. Each case aggregates combinations with the same fidelity contribution, and the probability of each case is the number of its constituent combinations divided by 16 (Table ~\ref{tab:grouped_cases}). This grouping aligns with Table ~\ref{table} in the main text and ensures no loss of statistical information.

\begin{table}[htbp]
\centering
\caption{Grouped Angle Combinations and Their Probabilities}
\label{tab:grouped_cases}
\begin{tabular}{ccccc}
\hline
\hline
Case & $c_1$ & $c_2$ & Number of Combinations & Probability \\
\hline
1 & 1 & 1 & 4 & 1/4 \\
2 & $\sqrt{2}/2$ & 0 & 6 & 3/8 \\
3 & 0 & $-1$ & 4 & 1/4 \\
4 & $-\sqrt{2}/2$ & 0 & 2 & 1/8 \\
\hline
\hline
\end{tabular}
\end{table}

\subsection{Calculation of the Average Fidelity for illegal user}

\textbf{Step 1: Fidelity Calculation for Each Case}

\bigskip
From Eq.~(\ref{15}) of the main text, the illegitimate-user fidelity is
\begin{align}
    F_{ill} &= \eta_0 k^2 (c_1 - 1)^2  \nonumber\\& + \frac{1}{4}\left[\eta_0 (\cosh 2r - \sinh 2r \cdot c_2)+ (1 - \eta_0)\right]
\end{align}

Substituting the $(c_1, c_2)$ pairs of each case into the above equation yields the fidelity for each group:

\textbf{Case 1:} $(c_1, c_2) = (1, 1)$
\begin{align}
    F_{ill}^{(1)} &= 0 + \frac{1}{4}[\eta_0 (\cosh 2r - \sinh 2r) + (1 - \eta_0)] \nonumber\\&= \frac{1}{4}[\eta_0 e^{-2r} + (1 - \eta_0)]
\end{align}

\textbf{Case 2:} $(c_1, c_2) = (\sqrt{2}/2, 0)$
\begin{align}
F_{ill}^{(2)} &= \eta_0 k^2 (\sqrt{2}/2 - 1)^2 + \frac{1}{4}[\eta_0 \cosh 2r + (1 - \eta_0)] \nonumber \\
&= \eta_0 k^2 \left(\frac{3}{2} - \sqrt{2}\right) + \frac{1}{4}[\eta_0 \cosh 2r + (1 - \eta_0)].
\end{align}

\textbf{Case 3:} $(c_1, c_2) = (0, -1)$
\begin{align}
    F_{ill}^{(3)} &= \eta_0 k^2 + \frac{1}{4}[\eta_0 (\cosh 2r + \sinh 2r) + (1 - \eta_0)] \nonumber\\
    &= \eta_0 k^2 + \frac{1}{4}[\eta_0 e^{2r} + (1 - \eta_0)]
\end{align}

\textbf{Case 4:} $(c_1, c_2) = (-\sqrt{2}/2, 0)$
\begin{align}
F_{ill}^{(4)} &= \eta_0 k^2 (-\sqrt{2}/2 - 1)^2 + \frac{1}{4}[\eta_0 \cosh 2r + (1 - \eta_0)] \nonumber \\
&= \eta_0 k^2 \left(\frac{3}{2} + \sqrt{2}\right) + \frac{1}{4}[\eta_0 \cosh 2r + (1 - \eta_0)]
\end{align}

\textbf{Step 2: Weighted Average of Illegitimate-User Fidelity}

\bigskip
The average illegitimate-user fidelity $F_{ill\_ave}$ is the weighted sum of $F_{ill}^{1}$ to $F_{ill}^{4}$, with weights equal to the probability of each case (Table~\ref{tab:grouped_cases}):
\begin{equation}
F_{ill\_ave} = \frac{1}{4}F_{ill}^{(1)} + \frac{3}{8}F_{ill}^{(2)} + \frac{1}{4}F_{ill}^{(3)} + \frac{1}{8}F_{ill}^{(4)}
\label{A8}
\end{equation}

\textbf{Coefficient of $\eta_0 k^2$:}
Substitute the $(c_1 - 1)^2$ terms of each case into Eq.~(\ref{A8}) and simplify:
\begin{align}
Coeff_{\eta_0 k^2} & = \left[ \frac{3}{8}\left(\frac{3}{2} - \sqrt{2}\right) + \frac{1}{4} + \frac{1}{8}\left(\frac{3}{2} + \sqrt{2}\right) \right]\nonumber\\
&= \frac{1}{4}(4 - \sqrt{2})\eta_0 k^2
\end{align}

\textbf{Second term (remaining part):} Substitute the hyperbolic function terms and constant terms into Eq.~(\ref{A8}), using $\cosh 2r = (e^{2r} + e^{-2r})/2$:
\begin{align}
&\frac{1}{4} \cdot \frac{1}{4}[\eta_0 e^{-2r} + (1 - \eta_0)] + \frac{3}{8} \cdot \frac{1}{4}[\eta_0 \cosh 2r + (1 - \eta_0)] \nonumber \\
&\quad + \frac{1}{4} \cdot \frac{1}{4}[\eta_0 e^{2r} + (1 - \eta_0)] + \frac{1}{8} \cdot \frac{1}{4}[\eta_0 \cosh 2r + (1 - \eta_0)] \nonumber \\
&= \frac{1}{4}[\eta_0 \cosh 2r + (1 - \eta_0)]
\end{align}

\textbf{Step 3: Final Expression}

\bigskip
Combining both terms, we obtain the average illegitimate-user fidelity:
\begin{equation}
\boxed{F_{ill\_ave} = \frac{1}{4}(4 - \sqrt{2})\eta_0 k^2 + \frac{1}{4}[\eta_0 \cosh 2r + (1 - \eta_0)]}
\end{equation}

This completes the derivation of Eq.~\ref{16} in the main text. The intermediate steps and combination factors are now fully documented for independent verification.

\section{Mapping Gaussian Key to Discrete Phase Angles}
\label{app:mapping}

This appendix details the procedure for mapping the Gaussian-distributed pre-shared key $k_0 \sim \mathcal{N}(0, k^2)$ to uniformly distributed discrete phase angles $\theta \in \Theta = \{0, \pi/2, \pi, 3\pi/2\}$, as used in Step~1 of the protocol.

\begin{figure*}[htbp]
  \centering
  \includegraphics[width=0.75\linewidth]{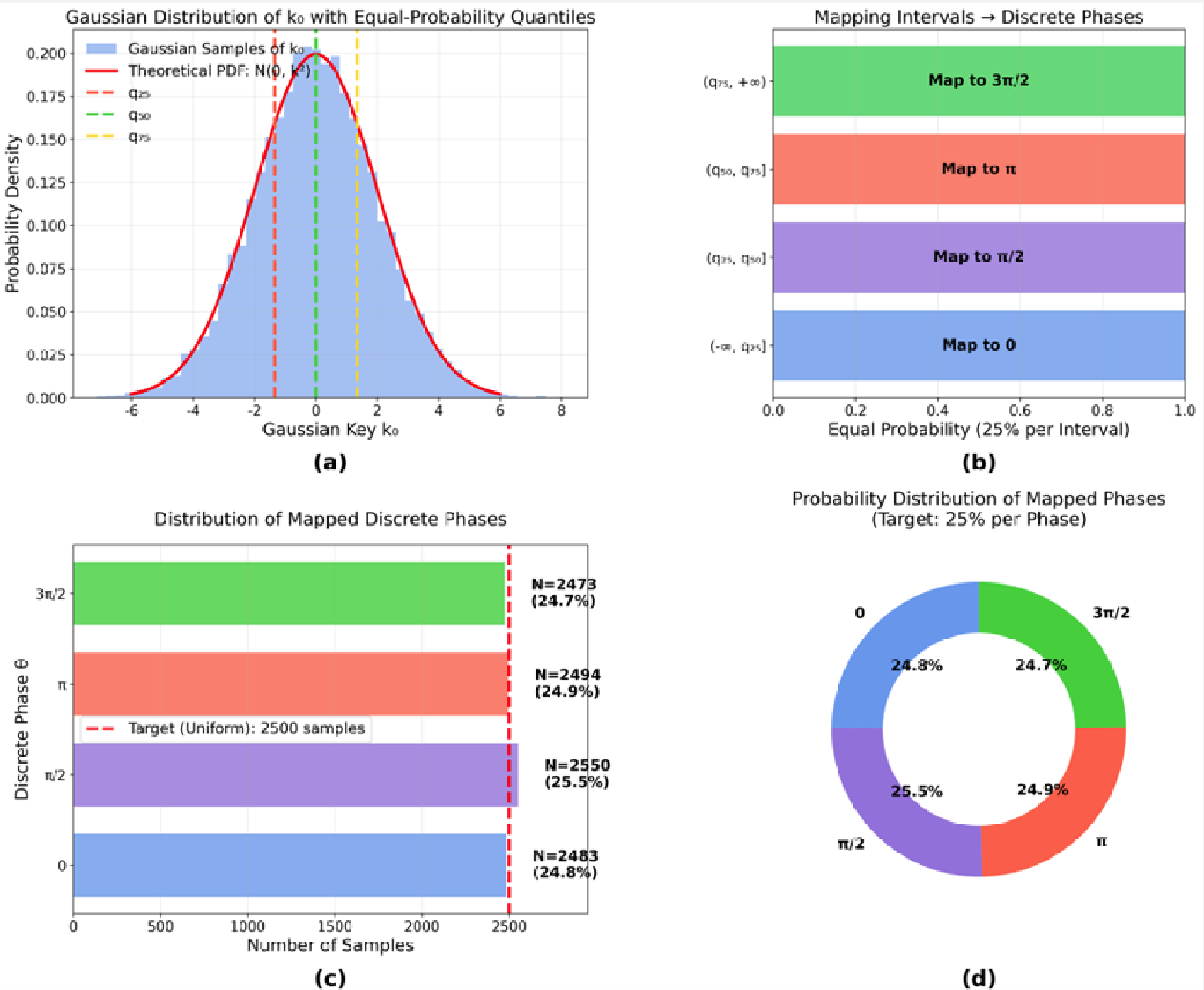}
  \caption{Mapping of Gaussian-distributed key $k_0 \sim \mathcal{N}(0, k^2)$ to discrete phase angles $\Theta = \{0, \pi/2, \pi, 3\pi/2\}$. (a) Gaussian distribution of $k_0$ with equal-probability quantiles $q_{25}$, $q_{50}$, $q_{75}$. (b) Mapping intervals to discrete phases with equal probability (25\% per interval). (c) Distribution of mapped discrete phases from Monte Carlo simulation ($N=10,000$ samples). (d) Probability distribution of mapped phases showing near-uniform distribution (target: 25\% per phase).}
  \label{fig:gaussian_to_phase_mapping}
\end{figure*}

To map the Gaussian-distributed pre-shared key $ k_0 \sim \mathcal{N}(0, k^2) $ to the discrete phase set $ \Theta = \{0, \frac{\pi}{2}, \pi, \frac{3\pi}{2}\} $ following a uniform distribution, the equal-probability mapping method is employed, which ensures each phase in $ \Theta $ has an identical probability of $25\%$. The detailed implementation process is described as follows:
\begin{itemize}
    \item \textbf{Step 1: Define Core Parameters}
        \begin{itemize}
        \item Gaussian key distribution: $ k_0 $ follows a Gaussian distribution with mean $ \mu = 0 $ and variance $ \sigma^2 = k^2 $ (i.e., $ k_0 \sim \mathcal{N}(0, k^2) $).
        \item Target discrete phase set: $ \Theta = \{0, \frac{\pi}{2}, \pi, \frac{3\pi}{2}\} $, containing $ K = 4 $ uniform discrete values.
        \item Target probability per phase: Since the total probability of the discrete uniform distribution sums to 1, each phase corresponds to a target probability of $ 1/K = 25\% $.
    \end{itemize}
    \item \textbf{Step 2: Calculate Equal-Probability Quantiles of the Gaussian Distribution}

    The Gaussian cumulative distribution function (CDF) $ F(x) = P(K_0 \leq x) $ and its inverse $ F^{-1}(p) $ (quantile function) are utilized to partition the Gaussian probability space into $ K = 4 $ intervals with equal probability ($25\%$ each):
    \begin{enumerate}[label=\Roman{enumi}.]
        \item Define quantile points corresponding to cumulative probabilities: $ p = [0.0, 0.25, 0.5, 0.75, 1.0] $ (5 points to form 4 intervals).
        \item Compute the quantiles of $ \mathcal{N}(0, k^2) $ using the inverse CDF:
        \begin{itemize}
            \item The quantile for $p=0.25$: $ q_{25} = F^{-1}(0.25) = -0.67449 \cdot k $
            \item The quantile for $p=0.5$: $ q_{50} = F^{-1}(0.5) = 0 $ (consistent with the Gaussian mean $ \mu = 0 $)
            \item The quantile for $p=0.75$: $ q_{75} = F^{-1}(0.75) = 0.67449 \cdot k $

        \end{itemize}
(Note: For the standard normal distribution $\mathcal{N}(0,1)$, $0.67449$ is the 75th percentile (i.e., $F^{-1}(0.75) \approx 0.67449$), and its negative value $-0.67449$ corresponds to the 25th percentile (i.e., $F^{-1}(0.25) \approx -0.67449$) due to the symmetry of the normal distribution. The 50th percentile (median) of $\mathcal{N}(0,1)$ is $0$ (i.e., $F^{-1}(0.5) = 0$), which aligns with the mean of the symmetric Gaussian distribution. Scaling these values by $k$ adapts them to the Gaussian key distribution $\mathcal{N}(0, k^2)$, ensuring the quantiles $q_{25}$, $q_{50}$, and $q_{75}$ partition the distribution into four equal-probability intervals.)
    \end{enumerate}

\item \textbf{Step 3: Map Gaussian Samples to Discrete Phases}

For each Gaussian key sample $ k_0 $, determine its corresponding interval based on the pre-computed quantiles and map it to the corresponding phase in $ \Theta $:
\begin{itemize}
    \item If $ k_0 \leq q_{25} $: Map to $ 0 $ (first phase, covering the leftmost $25\%$ probability of the Gaussian distribution).
    \item If $ q_{25} < k_0 \leq q_{50} $: Map to $ \frac{\pi}{2} $ (second phase, covering the next $25\%$ probability).
    \item If $ q_{50} < k_0 \leq q_{75} $: Map to $ \pi $ (third phase, covering the middle-right $25\%$ probability).
    \item If $ k_0 > q_{75} $: Map to $ \frac{3\pi}{2} $ (fourth phase, covering the rightmost $25\%$ probability).
\end{itemize}
The complete mapping process (including Gaussian distribution partitioning, interval-phase correspondence, and mapped phase distribution) is visually illustrated in Fig. \ref{fig:gaussian_to_phase_mapping}. The simulation results confirm that this method achieves approximately uniform distribution of discrete phases, which aligns with the design goal of the protocol.

\item \textbf{Step 4: Boundary Robustness Handling}

Extreme values of $ k_0 $ (e.g., $ k_0 \to \pm\infty $) are clipped to the boundary intervals to avoid index out-of-bounds issues, ensuring all samples are mapped to one of the four phases in $ \Theta $.

\end{itemize}

This method achieves strict uniform distribution of the phase parameter $ \theta $ by leveraging the quantile-based probability partitioning, with each phase occurring at a rate of approximately $25\%$, and it maintains good adaptability to the symmetric characteristic of the Gaussian distribution $ \mathcal{N}(0, k^2) $.


\nocite{*}
\bibliography{qia2025ref}

\end{document}